\documentclass[preprint]{aa}
\usepackage{natbib,graphicx}

\newcommand{\be}{\begin{equation}}
\newcommand{\ee}{\end{equation}}

\begin{document}

\date{\today}

\title{Monte Carlo radiative transfer \\
in molecular cloud cores}

\author{Jos\'e Gon\c{c}alves\inst{1,2}, Daniele Galli\inst{2}, \and Malcolm Walmsley\inst{2}}

\institute{Centro de Astronomia e Astrof\'{\i}sica da Universidade de Lisboa, Tapada da Ajuda, 1349-018 
Lisboa, Portugal
\and INAF-Osservatorio Astrofisico di Arcetri, Largo E. Fermi 5, I-50125 Firenze, Italy}

\offprints{J.~Gon\c{c}alves, \\
\email{goncalve@arcetri.astro.it}}

\date{Received / Accepted}

\authorrunning{Gon\c calves, Galli, Walmsley}
\titlerunning{Radiative transfer in cloud cores}

\abstract{We present the results of a three-dimensional Monte Carlo
radiative transfer code for starless molecular cloud cores heated by an
external isotropic or non-isotropic interstellar radiation field. The
code computes the dust temperature distribution inside model clouds with
specified but arbitrary density profiles. In particular we examine in
detail spherical (Bonnor-Ebert) clouds, axisymmetric and non-axisymmetric
toroids, and clouds heated by an external stellar source in addition to
the general interstellar field.  For these configurations, the code
also computes maps of the emergent intensity at different wavelengths and
arbitrary viewing angle, that can be compared directly with continuum
maps of prestellar cores.  In the approximation where the dust
temperature is independent of interactions with the gas and where the gas
is heated both by collisions with dust grains and ionization by cosmic
rays, the temperature distribution of the gas is also  calculated.
For cloud models with parameters typical of dense cores, the results
show that the dust temperature decreases monotonically from a maximum
value near the cloud's edge (14--15~K) to a minimum value at the cloud's
center (6--7~K).  Conversely, the gas temperature varies in a similar
range, but, due to efficient dust-gas coupling in the inner regions
and inefficient cosmic-ray heating in the outer regions, the gradient
is non-monotonic and the gas temperature reaches a maximum value at
intermediate radii.  The emission computed for these models (at
350~$\mu$m and 1.3~mm) shows that deviations from spherical
symmetry in the density and/or temperature distributions are generally
reduced in the simulated intensity maps (even without beam convolution),
especially at the longer wavelengths.
}

\maketitle

\section{Introduction}

Understanding the structure of pre--protostellar cores is an essential
step towards an understanding of protostellar evolution. The density
distribution immediately prior to the onset of gravitational collapse
defines the initial conditions for collapse and hence one has a strong
motivation to attempt to derive this density distribution from the
observations.  Since these objects are most commonly observed by means
of their millimeter dust emission, one has a strong interest in
understanding the temperature distribution of the dust in the
pre--protostellar core and its influence on the emergent intensity
distribution at millimeter-submillimeter wavelengths.  In this article,
we present a radiative transport code which has the interpretation of
mm-submm maps of pre--protostellar cores as its main goal.

The approach which we have adopted is influenced by the fact that, as a
rule, the density structure of these objects shows evidence for large
deviations from spherical symmetry.  More precisely, the maps of the
millimeter dust emission show clear departures from circular symmetry
(see e.g.  Andr\'e et al.~2000; Caselli et al.~2002a,b; Tafalla et al.
2002). In some cases there is also evidence for polarization
and hence for a magnetic field (Ward-Thompson
et al.~2000).  It seems likely to us that such behavior is caused by
magnetic fields of energy densities sufficiently large to influence the
core structure (see also Shu et al. 1987) and cause flattening along
field lines (Basu 2000, Jones \& Basu~2002).  We can attempt to
infer the density distribution on the basis of mm-submm continuum maps
if we can derive the temperature distribution in regions where
departures from spherical symmetry are important. A first attempt in
this direction was made by Zucconi et al.~(2001, hereafter ZWG, see
also Evans et al.~2001 for the spherically symmetric case) but these
authors neglected heating of dust grains due to re-absorption of
photons emitted by the cloud itself.  Here we present a Monte--Carlo
code without this limitation but which is capable of handling a variety
of geometries.  We expect that this type of analysis will be especially
useful as a tool for the interpretation of the high class data we
expect to come from future instruments like ALMA and HERSCHEL.

Some high quality data are already available and are consistent with a
picture in which the dust temperature of pre-protostellar cores
decreases towards the center (Ward-Thompson et al.~2002, Bianchi et
al.~2003).  ISO observations in the mid-IR (Bacmann et al.~2000) and in
the far-IR (Ward-Thompson et al.~2002) suggest dust temperature
gradients consistent with heating from the external interstellar
radiation field (ISRF). The available models predict a factor of $\sim
2$ increase in temperature from center to edge (ZWG, Evans et al.~2001,
Andr\'e et al.~2003), with the gradient dependent on the cloud
structure.  The predicted emission of several cores is in  reasonable
accord  with the data, and implies that the assumed equilibrium
configurations are either unstable or maintained by a magnetic field.

Additional observational constraints on the thermal structure of
prestellar cores are given by spatially resolved measurements of the
gas temperature from NH$_3$ observations. Low-mass cores show fairly
uniform gas temperature (e.g. Tafalla et al.~2002 for L1517B and L1498),
whereas massive quiescent cores in Orion show significant temperature
drops from edge to center (Li et al.~2003). Thus another check on our
understanding of core structure is possible if radiative transfer
models are also able to predict the gas temperature distribution resulting
from the balance of the relevant heating and cooling mechanisms.  We
therefore have also developed the capability to predict gas temperature
distributions for our model cores.

While techniques have been developed for the study of radiative transfer
in cold cores in one dimension (e.g. Leung~1976, Rowan-Robinson~1980,
Ivezic et al.~1997), their extension to three-dimensional geometries
is not straightforward. A Monte Carlo technique has the advantage that
it deals easily with any general geometry, independently of the density
distribution and the anisotropy of the incident radiation field. The
increasingly faster workstations available render this technique very
useful, and fully 3-D Monte Carlo radiative transfer codes have been
developed recently (e.g. Wolf et al.~1999, Niccolini et al.~2003).
However, in these latter studies the radiative sources are internal,
and this is a computationally distinct problem from a core heated from
the outside for two main reasons. First, as we are interested in cold
cores, it is reasonable to expect that they will be optically thin to
their own emission, and therefore, for methods that use iteration, the
temperature distribution quickly relaxes to the final value;
alternatively, for methods that follow individual packets until they
exit the domain, each individual packet will interact with the dust
often only once. Second, as packets are launched from the outside, a
wide range of spatial scales demands a prohibitively large photon
statistics, as the probability of a packet being launched in the
direction of the innermost cells becomes increasingly small.

In general, our paper confirms and supports the results recently obtained for
embedded prestellar cores by Stamatellos \& Whitworth~(2003) with an
independent but similar Monte Carlo radiative transfer code. While
Stamatellos \& Whitworth~(2003) focus on the important effect of the ambient
medium in which (spherical) cores are embedded, our paper mostly
emphasizes the consequences of deviations from spherical symmetry in
the density distribution and anisotropies of the ISRF. 
The two papers represent therefore complementary attempts to model
radiative transfer in molecular cloud cores in less idealized 
situations than those considered so far.

The structure of this paper is as follows. In Sect. 2, we set the
general problem of radiative transfer in a dusty cloud. In Sect. 3,
we describe the Monte Carlo method we have used. In Sect. 4, we
present our choice of opacities and interstellar radiation field and
briefly discuss their effect in the results. In esections 5 and 6 we
present our results for two- and three-dimensional models,
respectively. Finally, in Sect. 7 we summarize our conclusions.

\section{Radiative transfer in a dusty cloud}

In local thermodynamic equilibrium and ignoring the effects of
scattering of radiation by dust grains, the 
radiation field in a dusty cloud satisfies the equation of transfer
\be
{\hat{\bf n}}\cdot\nabla I_\nu({\bf r},{\hat{\bf n}})=
\rho({\bf r})\kappa_\nu({\bf r})
\{B_\nu[T_{\rm d}({\bf r})]- I_\nu({\bf r},{\hat{\bf n}})\},
\label{transfer}
\ee
where $I_\nu({\bf r},{\hat{\bf n}})$ is the specific intensity
of radiation (in Jy~ster$^{-1}$) at position ${\bf r}$ in the direction
${\hat{\bf n}}$ at frequency $\nu$, $\rho$ is the total (gas plus dust)
density, $\kappa_\nu$ the total absorption opacity, $T_{\rm d}$ is the
dust temperature, and $B_\nu$ is the Planck function,
\be
B_\nu(T_{\rm d})=\frac{2h \nu^3}{c^2}
\left[\exp\left(\frac{h\nu}{kT_{\rm d}}\right)-1\right]^{-1}.
\ee

The dust temperature distribution $T_{\rm d}({\bf r})$ is obtained by
solving the equation of balance of emitted and absorbed radiation,
\be
\int^\infty_0 \kappa_\nu B_\nu [T_{\rm d}({\bf r})]\;d\nu = 
\int^\infty_0 \kappa_\nu J_\nu({\bf r})\;d\nu,
\label{equilibrium}
\ee
where 
\be
J_\nu({\bf r})=\frac{1}{4\pi}\oint 
I_\nu({\bf r},{\hat{\bf n}})\;d\Omega
\ee
is the mean intensity of radiation (in Jy).

ZWG made the symplifying assumption that the dust grains are only
heated by the incident interstellar radiation field $J_\nu^{\rm
ISRF}$ (assumed isotropic), neglecting re-emission of radiation
absorbed by the grains themselves.  That is to say, they assumed that
the cloud is optically thin to its own radiation.  Under this
hypothesis, the equation of radiative transfer has the solution
\be
J_\nu({\bf r})=\frac{J_\nu^{\rm ISRF}}{4\pi}\oint 
\exp[-\tau_\nu({\bf r},{\hat{\bf n}})]\;d\Omega,
\label{approx}
\ee
where 
\be
\tau_\nu({\bf r},{\hat{\bf n}})=\int^{{\hat{\bf n}}R}_r 
\rho({\hat{\bf r}}r^\prime)\kappa_\nu({\hat{\bf r}}r^\prime)\;dr^\prime,
\ee
is the optical depth from a point ${\bf r}$ inside the cloud to a point
${\hat{\bf n}}R$ on the cloud boundary, in the direction ${\hat{\bf
n}}$. The calculation of the radiation field inside the cloud is thus
reduced to a large number of integrations along different directions
(rays) defined at each grid point. This simplification allowed ZWG to
model axisymmetric non-spherical density distributions, but has the
disadvantage of slightly underestimating the dust temperature,
particularly at the cold center of the cloud.

Our general approach is the following. First, for a given ISRF we
compute the radiation field at any point inside the cloud core and
solve the equation of balance of radiation emitted and absorbed by dust
grains to obtain the dust temperature distribution $T_{\rm d}({\bf
r})$. Second, we compute the flux emitted in a given view direction at
different wavelengths and we compare the results with observed spectral
energy distributions and monochromatic maps for specific
objects.

In several cases we have also considered  the effect of a spherical
envelope surrounding the cloud core on the resulting temperature
distribution.  The rationale for this is that these objects are always
embedded in a photon--dominated region (PDR) which transforms the
incident optical--UV field into mid and far IR radiation.  Often in
practice, this is radiation from transiently heated particles, either 
small grains or polycyclic aromatic hydrocarbons, emitting 
in the 3--30~$\mu$m range.
Since we wish to focus on the temperature structure of the high density
core interior, we have decided to assume that the radiation field
incident on embedded cores is cut off in the optical-UV wavelength
range and that the radiation absorbed in the PDR is reradiated in the
infrared.  We thus incorporate the photons reradiated by the PDR in the
external field.  The error involved here can  be estimated  by varying
the form of the ISRF and appears to be small (though see the discussion
of Andr\'e et al.~2003).  Our focus here is on the effects of the core
geometry upon temperature structure and we place minor emphasis on the
spectral energy distribution of the incident field.

We stress that our approach is general:  the density distribution, the
optical properties of the dust grains, the intensity and spectral
energy of the radiation source(s) are input parameters that can be
freely specified and can be adapted to model various astrophysical
situations.

\section{The Monte Carlo method}

In the Monte Carlo technique, the computational domain is divided in a
large number of cells of mass $m_i$ that absorb and emit radiation. The
energy that enters the computational domain is divided in $N_\gamma$
monochromatic packets of equal energy, that are launched stochastically
from the boundary and followed until they exit the cloud. These packets
may be absorbed by the dust particles, resulting in a temperature
increase of the cells where they are absorbed, and are immediately
reemitted to enforce radiative equilibrium. If the cell $i$ absorbs
$N_i$ packets, the energy absorbed by the cell per unit time is 
\be
\frac{dE_i^{\rm abs}}{dt}=\left(\frac{N_i}{N_\gamma}\right) L^{\rm
ISRF}, 
\ee 
where $L^{\rm ISRF}$ is the luminosity of the ISRF at the
cloud's surface, obtained integrating the mean intensity of the
(anisotropic) ISRF over the cloud's surface and over frequency, 
\be
L^{\rm ISRF}=\int_0^\infty d\nu \oint d\Omega \oint J^{\rm
ISRF}_\nu({\hat{\bf n}}){\hat{\bf n}}\cdot d{\bf S}.  
\ee 
Assuming LTE conditions, the energy emitted by the $i$-th cell per unit time is 
\be
\frac{dE_i^{\rm em}}{dt}=4\pi m_i\int_0^\infty \kappa_\nu 
B_\nu(T_{{\rm d},i})\,d\nu 
\label{luminosity} 
\ee 
where $T_{{\rm d},i}$ is the dust temperature of the $i$-th cell,
obtained by equating the absorbed and emitted energies per unit time,
\be 
\label{MC-equilibrium}
\int^\infty_0 \kappa_\nu B_\nu (T_{{\rm d},i})\;d\nu =
\left(\frac{N_i}{N_\gamma}\right)\frac{L^{\rm ISRF}}{4\pi m_i}.
\label{balance_mc} 
\ee 
The integral on the left-hand side of this equation can be tabulated
for a grid of values of the dust temperature and opacity. The solution
of Eq.~(\ref{balance_mc}) can then be easily obtained by iteration and
interpolation. In the limit to the continuum, $N_\gamma \rightarrow
\infty$ and $m_i \rightarrow 0$, Eq.~(\ref{balance_mc}) is equivalent
to Eq.~(\ref{equilibrium}), but differs from the approximation adopted
in ZWG, Eq.~(\ref{approx}), because in $N_i$ are included the packets
that may have been absorbed more than once in the cloud, and
subsequently re-emitted.

We handled re-radiation (and, therefore, energy conservation) using the
method devised by Bjorkman \& Wood (2001): once a packet is absorbed,
it is immediately re-emitted (to conserve energy) at a new frequency
determined by the local dust temperature. To achieve this, one starts
by noting that if a cell has emissivity $\kappa_\nu B_\nu (T_{{\rm
d},i} -\Delta T_{{\rm d},i})$ prior to absorbing a wave packet, then
after packet absorption its temperature increases by an amount $\Delta
T_{{\rm d},i}$, and the cell emissivity becomes $\kappa_\nu B_\nu
(T_{{\rm d},i})$. Thus, the increment in the cell emissivity is
\begin{eqnarray}
\Delta j_\nu & = & \kappa_\nu [B_\nu (T_{{\rm d},i}) 
- B_\nu (T_{{\rm d},i} - \Delta T_{{\rm d},i})] \nonumber \\
& & \simeq \kappa_\nu 
\left(\frac{dB_\nu}{dT_{\rm d}}\right)_{T=T_{{\rm d},i}}\Delta T_{{\rm d},i}.
\label{deltaem}
\end{eqnarray}
A photon packet is immediately reemitted with 
a frequency obtained from a probability distribution having the same
spectral shape as $\Delta j_\nu$, 
\be
\frac{dP_i}{d\nu}=\frac{\kappa_\nu}{K}
\left(\frac{dB_\nu}{dT_{\rm d}}\right)_{T=T_{{\rm d},i}},
\label{frequency_correction}
\ee
where $K=\int^\infty_0 \kappa_\nu (dB_\nu/dT_{\rm d})\; d\nu$ is a
normalization constant. The re-emitted photon packet is then followed
until a new interaction occurs, and the procedure is repeated until the
photon leaves the cloud. This method ensures energy conservation in a
statistical sense. Due to the low temperature of the cores, energy is
generally re-radiated at long wavelengths, to which the core has a low
optical depth, and the overall effect of re-emission on the dust
temperature is more significant in the core's cold interior but is
generally very small.

\subsection{Tests of the code}

We have tested our Monte Carlo code in different ways. Here we present
the results of two complementary tests, one checking the accuracy of 
the packet propagation procedure and one checking the equality
of emitted and absorbed energy.

\begin{figure}[ht]
\resizebox{8cm}{!}{\includegraphics{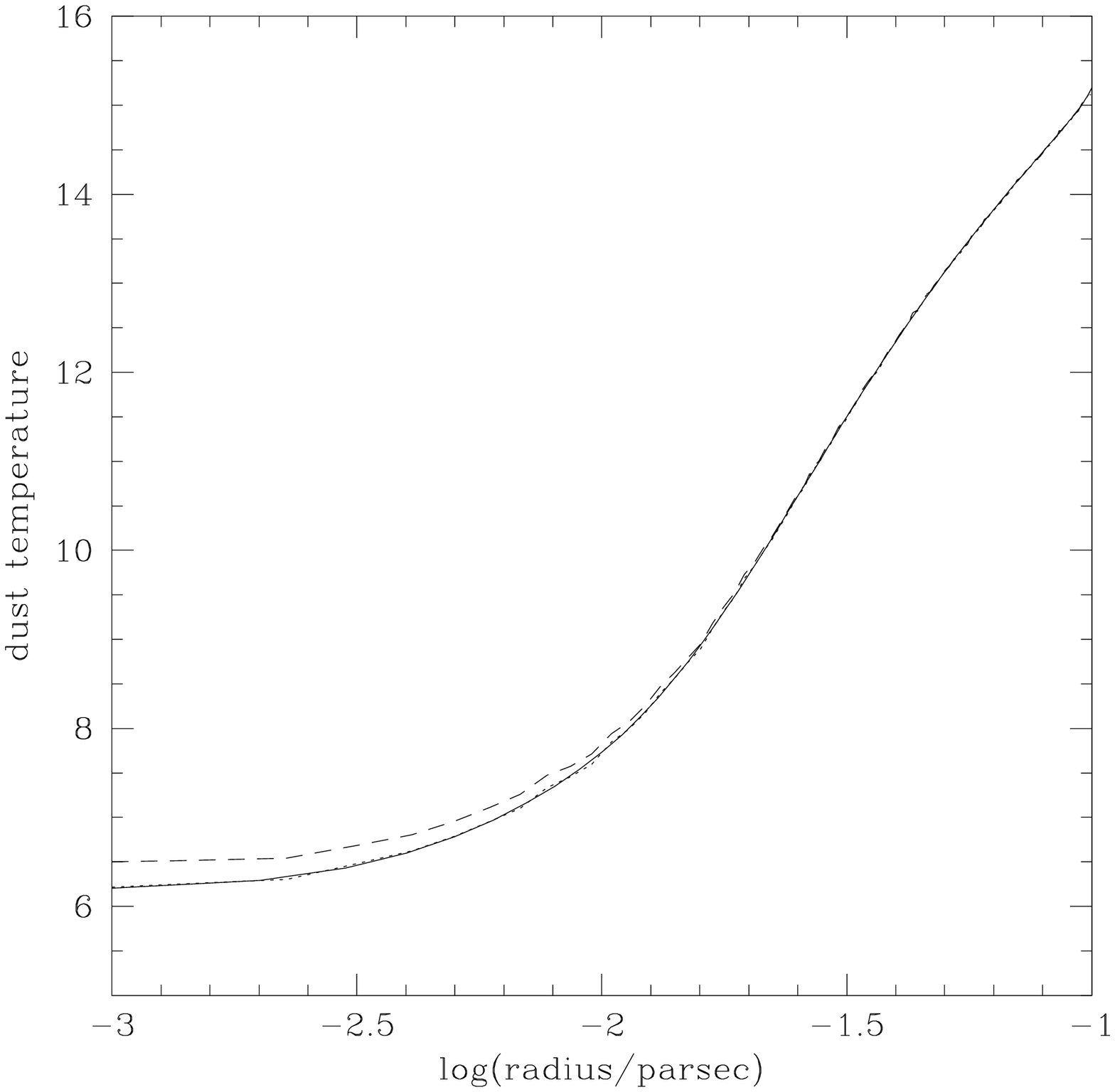}}
\caption{Comparison of dust temperature profiles obtained with
different methods for a BE sphere with central density $n({\rm
H}_2)=4.4\times 10^6$~cm$^{-3}$, gas temperature $T_g=12$~K, and radius
$R=0.1$~pc. The {\it solid} and {\it dotted curves} shows the result
obtained with the ray integration method of ZWG and the Monte Carlo
code, without including the effects of re-emission. The two results are
in excellent agreement. The {\it short-dashed curve} shows the dust
temperature computed with the Monte Carlo code including re-emission.
In all three cases the cloud has no envelope.}
\label{fig:test} 
\end{figure}

To test the packet propagation procedure, we have computed dust
temperatures profiles for spherically symmetric clouds, for which
results can also be obtained with the ray integration method of ZWG.
Specifically we have assumed the density profile of a Bonnor-Ebert (BE)
sphere, an isothermal, pressure-bounded, equilibrium model, that
reproduces the centrally flattened density distribution and the rapid
density decrease at large radii observed in most cloud cores.
Fig.~\ref{fig:test} shows the dust temperature as function of radius
for a BE with radius $R=0.1$~pc, central density $n({\rm H}_2)=4.4
\times 10^6$~cm$^{-3}$, and isothermal gas temperature $T_{\rm g}=12$~K
(note that the gas temperature here is  merely for the purpose of
specifying the assumed density distribution).  As shown in
Fig.~\ref{fig:test}, the result obtained using the ray integration
method (solid line), is identical to the Monte Carlo result using the
same approximation of optically thin re-emission (dotted line). We also
notice that when re-emission is taken into consideration (dashed line),
there is only a small increase of the central temperature, even in this
case, when the extinction through the core center is $A_{2.2\mu\rm{m}}
\simeq 25$, and the central dust temperature is little more than 6 K.

To test energy conservation, we have computed the energy emitted per
unit time by an arbitrary area element on the cloud's surface by
integrating $I_\nu$ (the solution of eq.~\ref{transfer}) and
$I_\nu^{\rm ISRF}$ over solid angles and frequencies, finding a
very small discrepancy between the two values (less than 1~\%).  This
test shows that for the given number and size of cells, the
accuracy of Eq.~(\ref{balance_mc}) and (\ref{frequency_correction}),
that contain approximations introduced by the discretization of the
computational volume, is very good.

\section {Dust opacities and the interstellar radiation field}

\subsection{The interstellar radiation field}

Following ZWG and Evans et al.~(2001), we adopt as a reference the ISRF
given by Black~(1994), which is an average for the solar
neighbourhood.  Scaling the intensity of the ISRF by a factor $G_0$
leads to a change in the dust temperature at all radii by a factor of
${G_0^{1/(4+\beta)}}$, where $\beta\simeq 2$ is the power law index for
the long wavelength opacity. 
Actually the radiation incident on molecular cloud cores
may differ not only in intensity but also in spectral shape, depending
on the degree of embedding of the cores in the ambient cloud and on the
possible vicinity to hot stars.  It appears for example to be the case
that many cores associated with the $\rho $~Oph star forming regions
are subjected to an incident ISRF roughly an order of magnitude larger
than the solar neighbourhood ISRF. An example of what can happen is
given by a recent study by Andr\'e et al. (2003) of a particular core,
heated by a ISRF higher than the Black field by an order of magnitude
in both the mid-IR and far-IR. This resulted in a higher central
temperature and a sharper gradient at the edge than in our models for a
similar density distribution.  Thus observers attempting to infer
density distributions from maps of the mm--submm dust emission must
bear in mind the fact that both the radiation field and the geometry
can play a role.

In this paper however, our objective is to study the effects of
geometry rather than those of the spectral characteristics or intensity
of the incident radiation field.  Geometry  in this context can
mean the geometry of the density distribution or that of the 
radiation field. We thus also consider a case where we use our Monte
Carlo model to consider an anisotropic incident ISRF as exists in many
galactic reflection nebulae.

\subsection{Dust opacities}

We adopt the dust opacities tabulated by Ossenkopf \& Henning~(1994,
hereafter OH) in the wavelength range 1~$\mu$m--1.3~mm. Recent work by
Bianchi et al.~(2003) and by Kramer et al.~(2003) has shown that these
theoretically derived opacities are consistent with the observed ratio
of mm dust continuum intensity and near infrared extinction.  It is
also clear from the study by OH that one can expect the dust opacity to
vary between the lower density outer layers of a core and the high
density interior where most of the heavy element content of the core is
in the form of ice of various sorts. We neglect such effects in this
study mainly because, as shown below, the variation between different
OH dust models is relatively mild.

We label the opacity according to the column number in Table~1 of OH:
OH1, standard MRN distribution; OH4, MRN with thin ice mantles; OH5,
MRN with thin ice mantles, after $10^5$~yr of coagulation at density
$10^6$~cm$^{-3}$; OH8, MRN with thick ice mantles, after $10^5$~yr of
coagulation at density $10^6$~cm$^{-3}$.

The OH opacities are tabulated for wavelengths between 1~$\mu$m and
1.3~mm. Within this range we have obtained values of the opacity at
arbitrary wavelengths by four-point interpolation and above 1.3~mm,
we have used a power law extrapolation based on the last two tabulated
values. We neglect scattering completely as we are  interested
in the transport of infrared photons.

\begin{figure}[ht]
\resizebox{8cm}{!}{\includegraphics{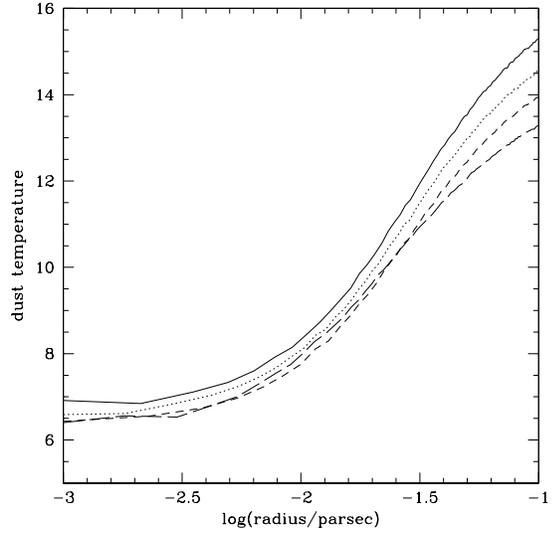}}
\caption{Dust temperature profiles (including re-emission) for the cloud
model shown in Fig.~1 for different choices of the dust opacity: {\it
solid curve}, OH1; {\it dotted curve}, OH4; {\it short-dashed curve},
OH5; {\it long-dashed curve}, OH8 (see text for the definitions). The
gas-to-dust ratio was varied  so that the total extinction at
2.2~$\mu$m is the same in all cases. The cloud is embedded in a spherical
outer envelope with total extinction $A_{2.2\mu\rm{m}}^{\rm env} = 0.1$.}
\label{fig:opacities} 
\end{figure}

The effect of the various opacities on the temperature profile of a
cloud is shown in Fig.~\ref{fig:opacities}, for the reference case of a
BE sphere with central density $n({\rm H}_2)=4.4\times 10^6$~cm$^{-3}$,
radius 0.1~pc, gas temperature $T_{\rm g}=12$~K, embedded in a
spherical envelope with total extinction $A^{\rm env}_{2.2\mu{\rm m}} =
0.1$.  In this particular example, the gas-to-dust ratio was varied for
each opacity prescription in order to keep fixed the total extinction
at 2.2~$\mu$m.

As Fig.~\ref{fig:opacities} shows, temperature variations due to
different choices of the opacity are generally small. The opacities
OH1, OH4 and OH5 have a similar dependence on wavelength, and the
resulting temperature difference is equivalent to scaling the intensity
of the ISRF by about $10\%$.  The OH8 opacity, having a slightly
different dependence on wavelength, results in a the temperature
profile slightly different from the other three cases. In all 
remaining models presented in this study we have used OH5.

\section{Two-dimensional models}

In this section we consider two particular situations that require a
two-dimensional modeling of the radiative transfer. In the first
example, the external radiation field is isotropic but the density
distribution is not spherically symmetric. In the second example, the
density profile is spherically symmetric but the incident radiation
field is anisotropic, as the cloud is heated by an external stellar
source.  Both situations are easily handled by our Monte Carlo code.

\subsection{Singular isothermal toroids}

As a first application of our Monte Carlo code, we consider the
radiative transfer in a singular isothermal toroid (Li \& Shu~1996), a
scale-free, axisymmetric equilibrium configuration of an isothermal
cloud under the influence of sef-gravity, gas pressure and magnetic
forces. These toroids are characterized by a single parameter, $H_0$,
representing the fractional amount of support provided by the magnetic
field. We choose the particular value $H_0=0.5$ to model a cloud with
moderate axial ratio ($\sim 2.5$) intermediate between a spherical
unmagnetized cloud and a magnetically dominated disklike
configuration.

It should be noticed that, in all models presented, we are simply
computing the dust temperature of pre-defined core density
distributions, and therefore the resulting core is not the solution of
the equilibrium equation if gas-dust coupling is assumed. Also, the
practical definition of the boundary as an isobaric surface is only
possible for spherical clouds, as in this case the isopycnic and
isobaric surfaces necessarily coincide. In all other geometries, we
take the boundary as the isopycnic surface corresponding to a 
specified value of the density.

\begin{figure}[ht]
\resizebox{8cm}{!}{\includegraphics[angle=270]{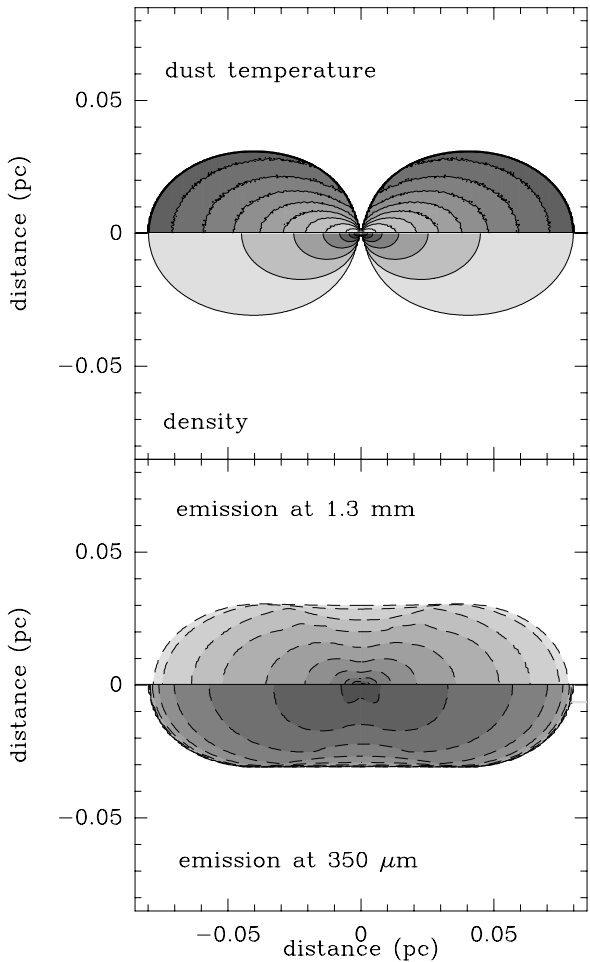}}
\caption{The {\it top panel} shows the dust temperature ({\it top
half}) and density distribution ({\it bottom half}) of a singular
isothermal toroid with $H_0=0.5$, seen edge-on. The density on the
boundary is $n({\rm H}_2)=10^4$~cm$^{-3}$ and the isopycnic curves are
logarithmically spaced by 0.5. The density distribution is azimuthally
symmetric and also symmetric with respect to the midplane. The maximum
outer temperature is 15~K and the isothermal curves are spaced by 1~K.
The {\it bottom panel} shows the emission at 1.3~mm ({\it top half})
and at 350~$\mu$m ({\it bottom half}). The isophotal curves are
logarithmically spaced by 0.2 starting from the lower value $0.01\times
I_\nu^{\rm max}$.  With the adopted opacity law (OH5), the peak values
are $I_\nu^{\rm max} =77$~MJy~sr$^{-1}$ at 1.3~mm and 382~MJy~sr$^{-1}$
at 350~$\mu$m (in this and in the following figures, the maps have not
been convolved with an observing beam).}
\label{fig:spit} 
\end{figure}

In Fig.~\ref{fig:spit} are shown both the density distribution of the
toroid and the resulting temperature distribution.  The boundary of
the configuration is defined as the isopycnic surface with $n({\rm
H}_2)=10^4$~cm$^{-3}$.  We choose this value of the density because
it represents the threshold value that characterizes typical dense
(``ammonia'') cores. For a density profile characterized by a $r^{-2}$
law, the visual extinction of the cloud material with density below this
threshold value corresponds to about $A_V\simeq 2$--3.  Notice that
the temperature varies over the boundary surface by a factor of about
two from the origin to the maximum radius, and this difference is due
to an increase of grain exposure to the ISRF as the radius increases,
an effect known also for disks of low aspect ratio (e.g. Spagna et
al.~1991), and is particularly enhanced in a toroid.  It is also evident
from Fig.~\ref{fig:spit} that the temperature decreases outward both in
the horizontal and vertical directions.

In Fig.~\ref{fig:spit}, we also show synthetic (unconvolved with an
observing beam) maps of the expected dust emission at 350~$\mu$m and
1.3~mm. In particular at 1.3~mm, the ``observed contours'' reflect
fairly faithfully the projected column density and hence the density
distribution. At 350~$\mu$m however, the intensity distribution is much
more extended than at 1.3~mm. The half-power dimensions are $\sim
1.5^{\prime\prime} \times 3.7^{\prime\prime}$ at 1.3~mm (in angular size
at a distance of 140 parsec)  but $18^{\prime\prime} \times
25^{\prime\prime}$ at 350~$\mu$m. Thus the effect of heating from the
exterior is mainly observable as an increase in size with frequency.

\subsection{Anisotropic Radiation Field}

While  the above discussion concerns cores of non-spherical geometry
subject to an isotropic incident radiation field, it should be noted
that one expects often to find cases where the incident radiation field
is anisotropic. Most obviously, this is the case in galactic reflection
nebulae where the radiation field from single early type stars may be
dominant. We therefore here (in a slight parenthesis to the discussion
elsewhere in this paper) demonstrate the application of our code to
such a case  and compute the expected temperature distribution in a
core subject to radiation from a nearby B star in addition to the
isotropic ISRF (note earlier work in this field by Natta et al. 1981).

The vicinity of a star to a prestellar core results in an anisotropic
radiation field, leading to a 2-D radiation transfer problem in the
case of a spherically symmetric core. It should be noted that this is a
case where the approximation of ZWG (Eq.~\ref{approx}) may not hold, as
the expected increase in dust temperature may make the core no longer
optically thin to its own radiation.

To study the effect of such a field, we start by considering a system
composed {\bf of a prestellar core represented as a BE sphere}, with
the physical parameters given in section 3.1, and a B3 star at a
distance of 0.15~pc from the edge of the core.  The large range of
spatial scales involved makes it difficult for a Monte Carlo program to
solve this problem consistently, by emitting packets from the star, and
following them through the ISM and the prestellar core. We avoid this
problem by assuming  that the stellar radiation is reprocessed by {\bf
a tenuous but spatially extended photodissociation region (PDR)
surrounding the core}, and use a fit to the spectral emission of the
PDR computed by D\'esert et al.~(1990). In this way, the situation
is reduced to the case treated in section 3.1, with the difference that
the incident radiation is characterized by a PDR spectrum, instead of
the standard ISRF, and a variable intensity over the surface of the
core. We further assume that the total luminosity over any area element on the
cloud's surface is the same as that obtained  assuming no attenuation
between the star and envelope.  Thus, the luminosity of a surface element
in spherical coordinates is 
\be 
dL(\theta) \propto \frac{1}{r^2(\theta)}
\cos\theta_{\rm i} \sin\theta \;d\theta, \label{Ltheta} 
\ee 
where ${\theta}$ is the angle between the point on the surface of the
core and the star as seen from the core's center, $r(\theta)$ is the
distance of that point to the star, and $\theta_{\rm i}$ is the angle
of incidence of the radiation. The proportionality constant is obtained
by setting the total incident luminosity equal to $(\Omega/4\pi)
L_\ast$, where $\Omega$ is the solid angle of the core as seen from the
star.

\begin{figure}[ht]
\resizebox{8cm}{!}{\includegraphics[angle=270]{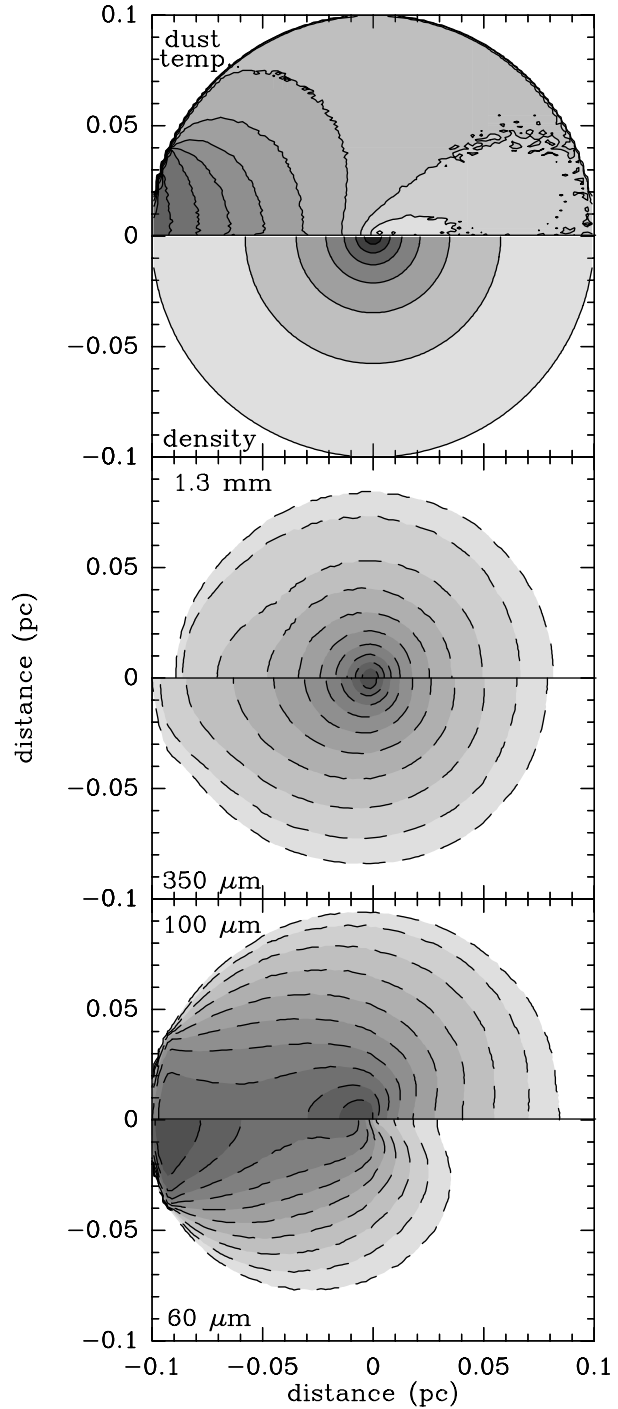}}
\caption{Model results for a spherical (Bonnor-Ebert) cloud core heated by the standard
ISRF and a B3 star located at 0.15~pc from the surface of the cloud (on
the left side in this figure). The core has central
density $n({\rm H}_2)=4.4\times 10^6$~cm$^{-3}$, gas temperature
$T_{\rm g}=10$~K, and radius $R=0.1$~pc.  The {\it top panel} shows the
dust temperature ({\it top half}\/) and the density ({\it bottom
half}\/). Isopycnic curves are logarithmically spaced by 0.5 and the
boundary density is $n({\rm H}_2)=2.3\times 10^3$~cm$^{-3}$. Isothermal
curves are spaced by 2~K starting from $T_{\rm d}=14$~K.  The {\it
middle} and {\it bottom panels} show the emission at 1.3~mm,
350~$\mu$m, 100~$\mu$m and 60~$\mu$m.  Isophotes are logarithmically
spaced by 0.2 starting from $0.01 \times I_\nu^{\rm max}$. The peak
intensity $I_\nu^{\rm max}$ is 78~MJy~sr$^{-1}$ at 1.3~mm,
3347~MJy~sr$^{-1}$ at 350~$\mu$m, 1404~MJy~sr$^{-1}$ at 100~$\mu$m and
200~MJy~sr$^{-1}$ at 60~$\mu$m.}
\label{fig:anis}
\end{figure}

The resulting temperature distribution is shown in the top panel of
Fig.~\ref{fig:anis}.  As expected, the cloud is hotter on the side facing
the star, and the temperature decreases both radially and azimuthally,
with the lowest temperature close to the centre of the core, which is
almost twice higher than the minimum temperature of the Black ISRF heated
core with the same density profile (Fig.~\ref{fig:test}). The ratio
$L_\ast/L^{\rm ISRF}$ is $\sim 20$, and the inverse square dependence
of $dL$ (see Eq.~\ref{Ltheta}) that makes this ratio over a surface
element be much larger close to the star, has the consequence that the
isotropic ISRF is negligible almost everywhere. The exception to this is
on the core's side opposite to the star, where the shielding provided
by the cloud's center suffices to make the isotropic ISRF the dominant
heating field.

In Fig.~\ref{fig:anis} we show the simulated emission maps at 1.3~mm,
350$~\mu$m, 100$~\mu$m and 60$~\mu$m. The 1.3~mm map shows slight
distortions due to the temperature  structure but clearly allows a
reasonable determination of the mass distribution in the BE sphere. As
one goes to shorter wavelengths however, the distortions become extreme
and at 60 microns, one sees essentially the heated surface of the
cloud. Thus maps at different wavelengths may allow the identification
of a heating source and show where the star is along the line of sight
relative to the cloud. In such situations, there will often be mid--IR
data available which will additionally define the geometry of the
``envelope'' or PDR layer hypothesised in our analysis.

\section{Three-dimensional models}

Reverting to the case of an isotropic radiation field,
we now consider the radiative transfer in a fully 3-D density distribution 
built on the results of
Galli et al.~(2001) for magnetized disks with uniform mass-to-flux
ratio $M(\Phi)/\Phi$ (isopedic).  A computationally convenient
characteristic of these models is the existence of non-axisymmetric but
analytical solutions of the equations of magnetostatics, corresponding
to surface density distributions $\Sigma$ separable in polar coordinates
$(\varpi,\varphi)$:
\be
\Sigma(\varpi,\varphi)=\frac{\Theta a^2}{2\pi \epsilon G\varpi}S(\varphi),
\label{SID}
\ee
where 
\be
S(\varphi)=\frac{\sqrt{1-e^2}}{1+e\cos\varphi},
\label{angularpart}
\ee
with $0<e<1$. Here, $\Theta\ge 1$ and $\epsilon\le 1$ represent the
increase of gas pressure and the reduction of the gravitational constant
due to magnetic pressure and magnetic tension, respectively, related to
the nondimensional mass-to-flux ratio $\lambda=2\pi G^{1/2}M(\Phi)/\Phi$
by the relations (Shu \& Li~1997) 
\be 
\epsilon=1-\frac{1}{\lambda^2 },
\qquad \Theta = \frac{\lambda^2+3}{\lambda^2+1}.  
\label{magnetic} 
\ee

We construct, in a rather {\it ad hoc} way, a 3-D model assuming a
vertical stratification of gas in hydrostatic equilibrium and imposing the
condition that the column density of the resulting density distribution
reproduces Eq.~(\ref{SID}), obtaining
\be
\rho(\varpi,\varphi,z)=\frac{a^2 \Theta^2}{8\pi \epsilon^2 G\varpi^2}
S^2(\varphi)\;{\rm sech}^2\left[\frac{z}{H(\varpi,\varphi)}\right],
\label{SID3d}
\ee
where $H(\varpi,\varphi)=a^2/\pi G\Sigma(\varpi,\varphi)$.

As a specific example, we adopt the parameters derived by Galli
et al.~(2001) to match the thermal dust emission map obtained by
Ward-Thompson et al.~(1999) for the starless core L1544. We assume an
effective sound speed $a= 0.21$~km~s$^{-1}$, a nondimensional mass-to-flux
ratio $\lambda=2$, an eccentricity $e=0.54$, and we define the core
boundary by the isopycnic surface $n({\rm H}_2)=10^4$~cm$^{-3}$.

\begin{figure*}[t]
\resizebox{15cm}{!}{\includegraphics[angle=270]{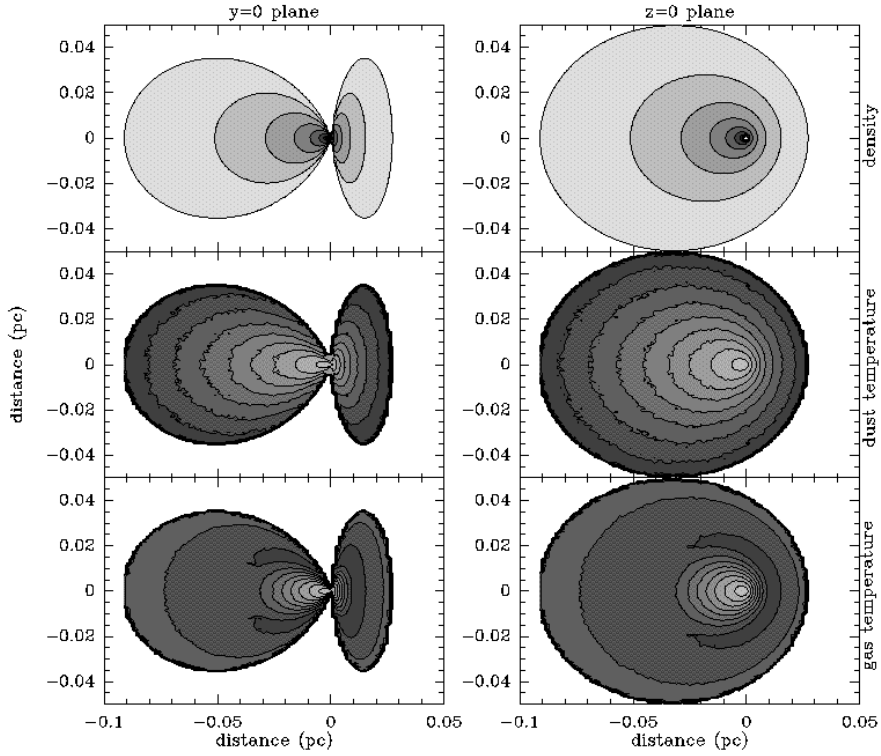}}
\caption{Density and temperature distribution in the magnetic 3-D model. Panels
on the left (right) side show results in $y=0$ ($z=0$) plane, respectively.
The {\it top panels} show the gas density, with isopycnic curves logarithmically
spaced by 0.5, and density on the
boundary $n({\rm H}_2)=10^4$~cm$^{-3}$. The {\it middle panels} show
the dust temperature distribution, with isothermal curves spaced by 1~K
between the minimum (7~K) and maximum (14~K) values. The {\it bottom
panels} show the gas temperature, with isothermal levels ranging from 7
to 14~K. }
\label{fig:sid} 
\end{figure*}

For this model we have also computed the gas temperature. To do this,
we first notice that the energy deposited in the gas by cosmic ray
ionization is negligible when compared to the energy absorbed by the dust,
so that the energy transfer between  gas and dust will not significantly
affect the grain temperature. Therefore, it is only necessary to solve
the equation of thermal equilibrium of the gas,
\be
\Gamma_{\rm cr}-\Lambda_{\rm g}-\Lambda_{\rm gd}=0, 
\ee
with ${\Gamma_{\rm cr}}$ the cosmic-ray heating rate, ${\Lambda_{\rm
g}}$ the gas cooling rate by molecular and atomic transitions, and
${\Lambda_{\rm gd}}$ the gas-dust energy transfer rate. For these
quantities, we assume the values given by Goldsmith~(2001). We also assume
a CO depletion factor of the form $f_{\rm dep}=\exp(n/n_{\rm dep})$,
where the critical density for CO depletion is taken to be $n_{\rm
dep}=5.5\times 10^4$~cm$^{-3}$.  Fig.~\ref{fig:sid} shows the density,
dust temperature and gas temperature of our model, with the left 
and right panels showing the $y=0$ and the $z=0$ plane, respectively.

This 3-D model and the 2-D toroidal model presented in the previous
section show qualitatively similar temperature distributions, but the
mismatch between isopycnic and dust isothermal surfaces is enhanced in
the 3-D model, particularly in the $y=0$ plane where the geometry is
more complex.

The gas temperature, as it depends on both the dust temperature
and on the gas density, shows a different distribution from either
(Fig.~\ref{fig:sid}). At the dense center, the temperatures of the
gas and the dust are well coupled, but they are decoupled
at the lower density boundary. This results in a non-monotonic behaviour 
of the gas temperature, which reaches a maximum at intermediate density 
(an effect that also occurs in 1D geometry, as shown in
Galli et al.~2002). It is also evident, from Fig.~(\ref{fig:sid})
that the gas temperature has a slightly higher value in the right side of
the singularity than at the left, which is a consequence of the different
dust temperature values at the density where the turnover in the 
gas temperature occurs on the two sides of the singularity.

\begin{figure}[ht]
\resizebox{8cm}{!}{\includegraphics[angle=270]{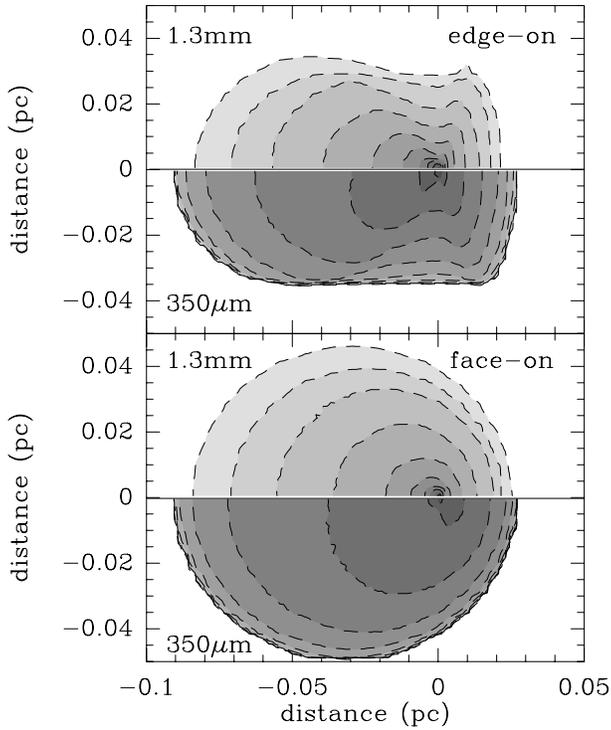}}
\caption{Normalized emission for the 3-D model, seen edge-on ({\it
top panel}\/) and face-on ({\it bottom panel}\/). Each panel shows the
emission at 1.3~mm ({\it top half}\/) and at 350~$\mu$m ({\it bottom
half}\/). Isophotal curves are logarithmically spaced by 0.2 starting from
the lower value $0.01\times I_\nu^{\rm max}$.  With the adopted opacity
law (OH5), the peak values are: $I_\nu^{\rm max}=89$~MJy~sr$^{-1}$ at
1.3~mm and 460~MJy~sr$^{-1}$ at 350~$\mu$m (edge-on case); $I_\nu^{\rm
max} =72$~MJy~sr$^{-1}$ at 1.3~mm and 307~MJy~sr$^{-1}$ at 350~$\mu$m
(face-on case).}
\label{fig:sidem}
\end{figure}

Fig.~\ref{fig:sidem} shows simulated dust emission maps at 1.3~mm and
350~$\mu$m for the side view and top view. The flattening towards the
center due to the dust temperature gradient is evident from the
comparison between the emission at the two wavelengths. The vertical
density stratification has a fairly strong effect on the shape of the
map seen from the side, but is not noticable from the top. In general,
one notes  a slight ``bottleneck'' deviation from a purely elliptical
shape of the singular isothermal disk model present in Galli et
al.~(2001).  We note also that the ``cometary shape'' of the edge-on
1.3~mm map is rather similar to that observed  towards L1544 and  L63
by Ward-Thompson et al.~(1999). 

\section{Conclusions}

In this paper we have presented a Monte Carlo model for radiative
transfer applicable to the study of the physical conditions in starless
molecular cloud cores. The method described is able to compute accurately
and efficiently the dust and gas temperature distribution in clouds of
specified but arbitrary density profiles heated by the ISRF, an external
stellar source, or a combination thereof.  In addition, the method allows
the computation of synthetic maps of the emission at near-infrared and
submillimeter wavelengths for arbitrary viewing angles, and to predict
the emergent spectral energy distribution.  We have successfully tested
our code by detailed comparison with one-dimensional calculations in
spherical symmetry (ZWG, Galli et al.  2002) and have shown that the
assumption of effective optical thinness is a good one for prestellar
cores.  As preliminary applications of our code, we have considered the
radiative transfer in model clouds of increasing degrees of asymmetry
and in a spherical cloud core heated by a neighboring star.

With our method, we confirm the results of ZWG, Evans et al.~(2001),
and Stamatellos \& Whitworth~(2003) that the dust temperature in cloud
models with physical characteristics typical of dense cores varies
monotonically between a minimum value at the center (6--7~K) and a
maximum value near the cloud's edge (14--15~K) for the standard ISRF.

In general, maps at submillimeter wavelengths closely reproduce the
column density distribution, even for the most non-isothermal
configurations. In the Rayleigh-Jeans regime, in fact, temperature
differences of a factor $\sim 2$ along the line-of-sight are washed out
by the much larger variation of the density.  However, the decrease of
the dust temperature at the cloud center has the consequence that
submillimeter maps sample preferentially the low-density envelope of
prestellar cores, and are therefore unable to distinguish centrally
peaked (pivotal) configurations from more flattened density profiles.
We also note  that our three-dimensional model based on the work
of Galli et al.~(2001) produces in natural fashion the cometary
structure observed towards many prestellar cores.

For a spherical core heated by an external stellar source close to
the cloud's surface, the temperature distribution is non-spherical, but
the submillimeter emission generally reproduces the core's column
density profile.  Deviations from circular symmetry become extremely
apparent on the other hand shortward of 100~$\mu$m where one samples
mainly the heated surface layer.

\begin{acknowledgements}
We thank the referee S. P. Goodwin for useful comments and A. Natta
for constructive criticism.  We acknowledge financial support from the
EC Research Training Network ``The Formation and Evolution of Young
Stellar Clusters'' (HPRN-CT-2000-00155).  JG acknowledges support from
the scholarship SFRH/BD/6108/2001 awarded by the Funda\c c\~ao para a
Ci\^encia e Tecnologia (Portugal).  \end{acknowledgements}

\bibliography{}

\bibliographystyle{aa}

\clearpage

\clearpage

\clearpage

\end{document}